\documentclass[twocolumn,showpacs,preprintnumbers,amsmath,amssymb,superscriptaddress,prl,floatfix]{revtex4}

\pdfoutput=1

\usepackage{graphicx}
\usepackage{dcolumn}
\usepackage{bm}
\usepackage{epsfig}
\usepackage[american]{babel}
\usepackage{color}
\usepackage{textcomp}
\usepackage{amsmath}
\usepackage{amssymb}
\usepackage{amsfonts}

\begin{document}

\title{Phonon--mediated non--equilibrium interaction between nanoscale devices}

\author{G.\,J.\ Schinner}
\affiliation{Center for NanoScience and Fakult\"at f\"ur Physik, Ludwig-Maximilians-Universit\"at, Geschwister-Scholl-Platz 1, 80539 M\"unchen, Germany}

\author{H.\,P.\ Tranitz}
\affiliation{Institut f\"ur Experimentelle und Angewandte Physik,Universit\"at Regensburg, 93040 Regensburg, Germany}

\author{W. Wegscheider}
\affiliation{Institut f\"ur Experimentelle und Angewandte Physik,Universit\"at Regensburg, 93040 Regensburg, Germany}

\author{J.\,P.\ Kotthaus}
\affiliation{Center for NanoScience and Fakult\"at f\"ur Physik, Ludwig-Maximilians-Universit\"at, Geschwister-Scholl-Platz 1, 80539 M\"unchen, Germany}

\author{S.\ Ludwig}
\affiliation{Center for NanoScience and Fakult\"at f\"ur Physik, Ludwig-Maximilians-Universit\"at, Geschwister-Scholl-Platz 1, 80539 M\"unchen, Germany}

\begin{abstract}
Interactions between mesoscopic devices induced by interface acoustic phonons propagating in the plane of a two--dimensional electron system (2DES) are investigated by phonon--spectroscopy. In our experiments ballistic electrons injected from a biased quantum point contact emit phonons and a portion of them are reabsorbed exciting electrons in a nearby degenerate 2DES. We perform energy spectroscopy on these excited electrons employing a tunable electrostatic barrier in an electrically separate and unbiased detector circuit. The transferred energy is found to be bounded by a maximum value corresponding to Fermi--level electrons excited and back--scattered by absorbing interface phonons. Our results imply that phonon--mediated interaction plays an important role for the decoherence of solid--state--based quantum circuits.
\end{abstract}


%03.67.-a Quantum information, 
%68.65.-k Low-dimensional, mesoscopic, and nanoscale systems: structure and nonelectronic %properties 
%73.23.-b Electronic transport in mesoscopic systems, 
%73.63.-b Electronic transport in nanoscale materials and structures

\maketitle 

Nanoscale electronic circuits dominate present information technologies. Based on their coherent dynamics they are also considered as candidates for future quantum information processing \cite{Petta05,Koppens06}. Therefore, it is important to understand and control decoherence--inducing processes, such as the non--equilibrium back--action of a biased quantum point contact (QPC), widely used as single electron detector. However, the details of the relevant back--action mechanisms are not yet understood and a matter of ongoing investigations \cite{Kouwenhoven06,Khrapai06,Ensslin07,Gustavson08,Taubert08}. 

Phonon--induced currents in a two--dimensional electron system (2DES) have been evidenced in thermopower experiments \cite{Fletcher88,Ruf88} and also directly imaged with ballistically injected phonons \cite{Karl88}. In our experiments we employ a spectrometer, conceptually similar to a so--called lateral tunneling hot--electron amplifier \cite{Palevski89}, to analyse the energy of excited electrons in a 2DES and to study energy transfer mechanisms between mesoscopic circuits.

\begin{figure}[ht]
\centering
\includegraphics[angle=0]{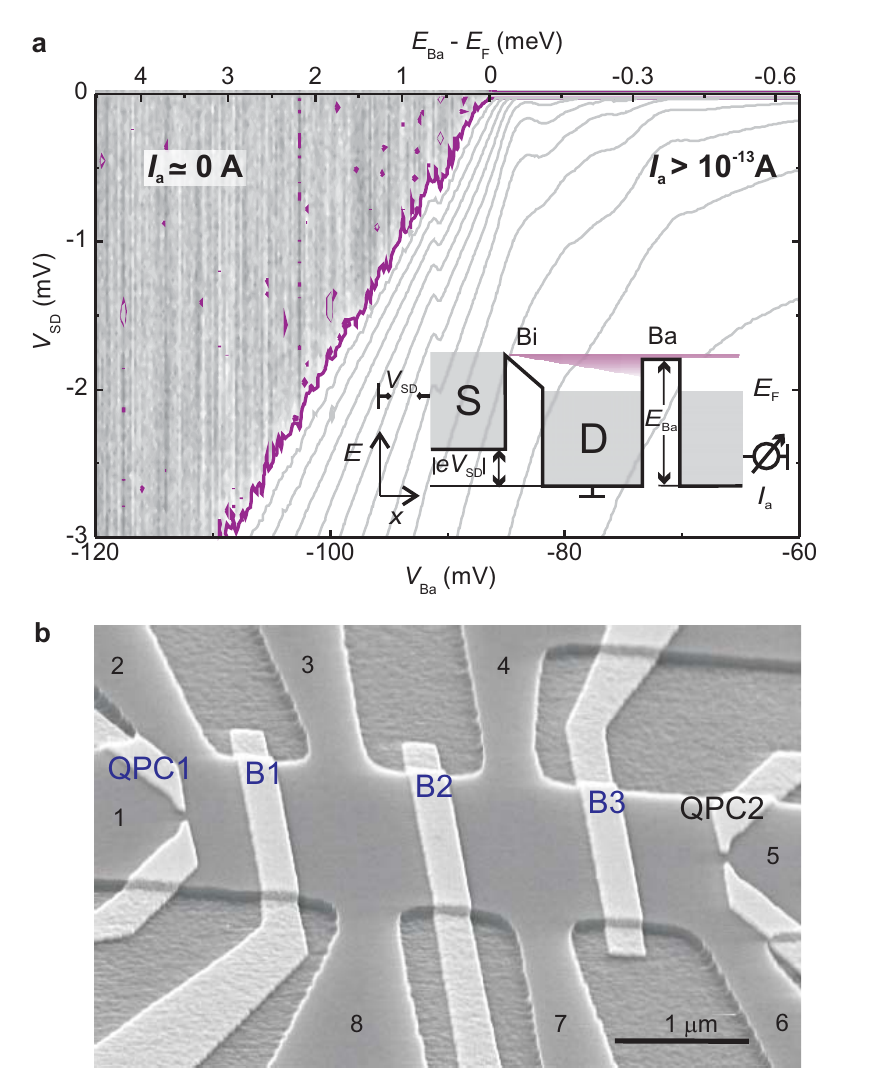}
\caption{\label{fig1}
\textbf{(a) Barrier calibration:} (color online) Inset: calibration setup (for details see main text). Main figure: Current $I_\mathrm{a}$ (gray scale for $I_\mathrm a<100\,$fA,  contour lines of constant current spaced by a factor of $3.3$ for $I_\mathrm a>100\,$fA) as a function of $V_\mathrm{SD}$ and the gate voltage $V_\mathrm{Ba}$. The current onset ($I_\mathrm{a}\simeq100\,$fA), highlighted in purple, serves as calibration. The resulting energy scale is displayed on the top axis. \textbf{(b) Sample geometry:} A Hall--bar (dark gray) with eight Ohmic contacts ($1,\,2,\,\dots,\,8$) is shaped from a GaAs/AlGaAs--heterostructure using electron--beam lithography (scanning electron micrograph). The Hall--bar is partly covered by metal gates (light gray) used to electrostatically define potential barriers (B1, B2, B3) and quantum point contacts (QPC1, QPC2).}
\end{figure}

The inset of Fig.\ \ref{fig1}\textbf{a}
sketches the calibration procedure of the energetic height $E_\mathrm{Ba}$ of an analyzer barrier Ba to be employed for quantitative energy spectroscopy. Hot electrons, injected across a barrier Bi into a degenerate Fermi--sea of cold electrons, move ballistically with an excess kinetic energy of $E_{\mathrm{kin}}-E_\mathrm{F}\le |e V_\mathrm{SD}|$ towards Ba. As long as $E_\mathrm{Ba}<E_{\mathrm{kin}}$ some of these electrons pass Ba resulting in an analyzer current $I_\mathrm a$, while $I_\mathrm a$ vanishes for $E_\mathrm{Ba}>E_{\mathrm{kin}}$. The onset of $I_\mathrm a (V_\mathrm{SD})$ at $E_\mathrm{Ba}=E_\mathrm{F}+|e V_\mathrm{SD}|$ serves as calibration of the barrier height $E_\mathrm{Ba}$. The result of such a calibration measurement is plotted in Fig.\ \ref{fig1}\textbf{a} displaying $I_\mathrm a$ (gray scale and contour lines) as a function of the gate voltage $V_\mathrm{Ba}$ and the bias $V_\mathrm{SD}$. The ballistic motion of the electrons insures a straight line of current onset (purple), converting the gate voltage $V_\mathrm{Ba}$ (bottom scale of Fig.\ \ref{fig1}\textbf{a}) to the barrier height $E_\mathrm{Ba}$ (top scale). For $E_\mathrm{Ba}<E_{\mathrm{F}}$ a calibration is obtained by utilizing quantization of the electronic density of states into Landau levels with well known energies in a perpendicular magnetic field \cite{haug88,komiy89}. Note that one calibration point, namely for $E_\mathrm{Ba}=E_{\mathrm{F}}$, is in addition obtained by applying a voltage across Ba and measuring the linear response current. Importantly, at $E_\mathrm{Ba}=E_{\mathrm{F}}$ all three calibration methods are consistent.

A scanning electron micrograph of our spectrometer is pictured in Fig.\ \ref{fig1}\textbf{b}. It is a mesoscopic Hall--bar shaped by wet--etching from a GaAs/AlGaAs heterostructure. The Hall--bar contains 90\,nm below the surface a 2DES with a Fermi energy of $E_{\mathrm{F}}\simeq14\,$meV and an electron elastic mean free path of $l_{\mathrm{m}}\simeq14\,\mathrm{\mu}\mathrm{m}$. Three 300\,nm wide top gates (light gray in Fig.\ \ref{fig1}\textbf{b}) are designed to cross the entire Hall--bar. By applying negative voltages to these gates, tunable potential barriers (B1, B2, and B3), completely suppressing tunneling, can be realized \cite{comment1}. In addition, at each end of the Hall--bar a QPC can be electrostatically defined by a pair of top gates. All experiments are performed in a dilution refrigerator at a base temperature of $T_{\rm bath}=20$\,mK.

\begin{figure}
\begin{center}
\includegraphics[width=89mm]{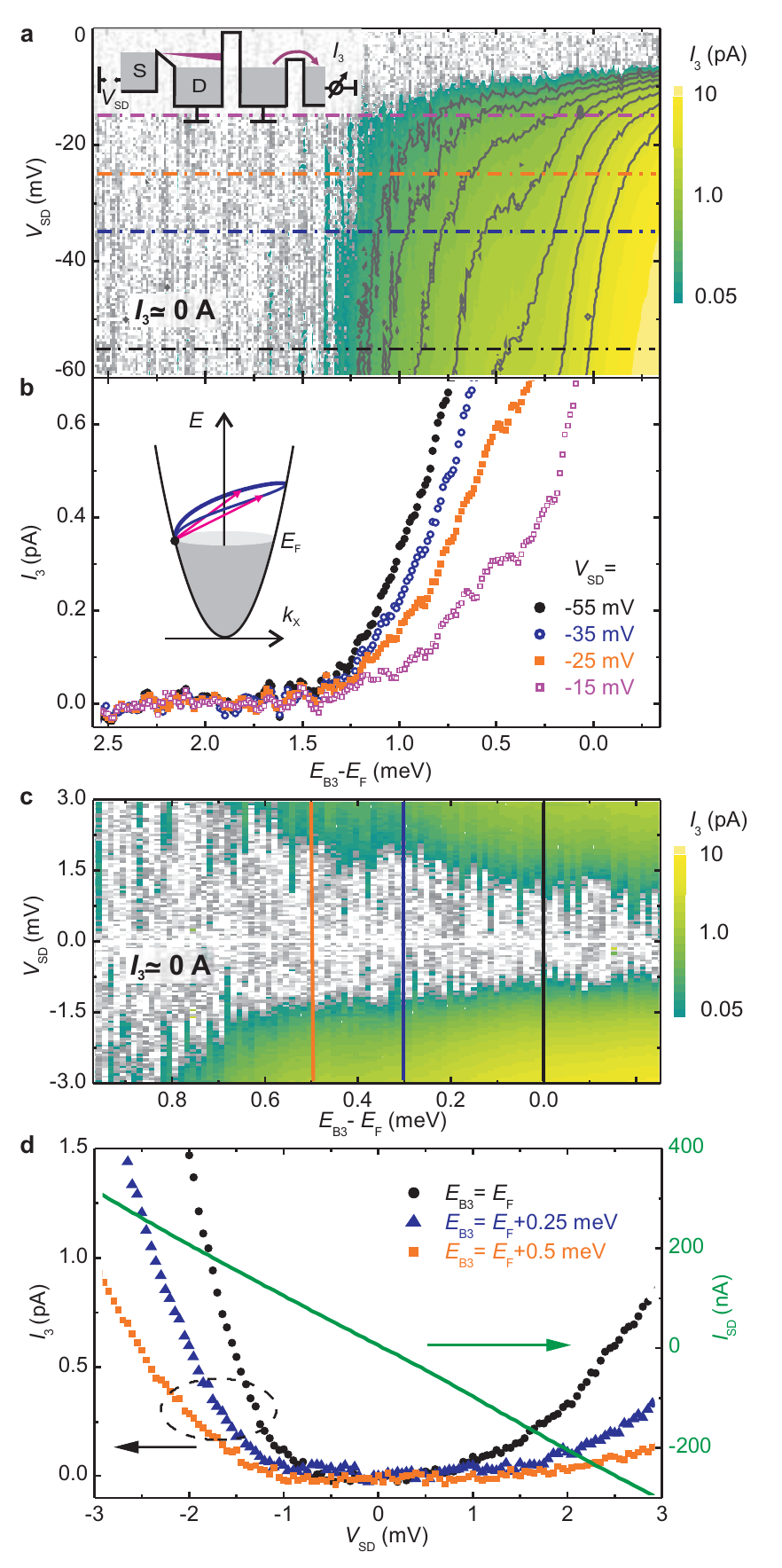}
\vspace{-5mm}
\caption{\label{fig2} \textbf{Phonon--driven current:} (color online)
\textbf{(a)} Analyzer current $I_3=I_{\mathrm a}$ (gray scale, color for $I_3\ge50\,$fA) across B3 (as analyzer barrier Ba) as a function of its energetic height $E_\mathrm{B3}-E_{\mathrm F}$ in the bias range $0\ge V_\mathrm{SD}\ge -60\,$mV applied across QPC1 (as emitter). Barrier B1 is opaque for electrons and separates emitter and detector as sketched in the inset. B2 is left open ($E_\mathrm{B2}\ll E_\mathrm{F}$). Contour lines of constant current are spaced by a factor of $1.7$.
\textbf{(b)} $I_3$--$E_\mathrm{B3}$ traces along the horizontal lines in \textbf{a}. The inset sketches relevant phonon absorption processes for an electron at the Fermi--level $E_\mathrm{F}$.
\textbf{(c)} Analyzer current $I_3$ in the bias range $3\,\mathrm{mV}\ge V_\mathrm{SD}\ge -3\,$mV. Currents comparable to those at larger bias are achieved by lowering the injector barrier resistance. For the detailed configuration see main text.
\textbf{(d)} $I_3$--$V_\mathrm{SD}$ traces along the vertical lines in \textbf{c}. Also plotted is the injector current $I_\mathrm{SD}$ (rhs--axis).}
\end{center}
\end{figure}

To spectroscopically study the energy transfer mechanisms between two adjacant mesoscopic devices we bias one of the barriers (B1) with a large negative voltage. As a result B1 is opaque for electrons and electrically separates the driven injector circuit from an unbiased detector circuit. As sketched in Fig.\ \ref{fig2}\textbf{a},
hot electrons injected across a QPC (QPC1) move ballistically along the Hall--bar until they are reflected at barrier B1 and eventually leave the Hall--bar via a grounded side--contact. In the detector circuit barrier B2 is left open for electrons ($E_\mathrm{B2}\ll E_\mathrm{F}$) and B3 is used as analyzer barrier.

Although the detector circuit is unbiased we observe a current $I_3$ across the analyzer barrier B3. Hence, energy is transmitted across B1 while electrons are always reflected. In Fig.\ \ref{fig2}\textbf{a} the measured $I_3$ is displayed for a large bias regime $-60\,\mathrm{mV}\le V_\mathrm{SD} \le 0$ and as a function of the excess barrier height $E_\mathrm{B3}-E_\mathrm{F}$. Strikingly, even at a large energy of injected electrons $\left|eV_\mathrm{SD}\right|=60\,\mathrm{meV}$, the analyzer current vanishes whenever the analyzer barrier height exceeds $E_\mathrm{B3}\simeq E_\mathrm{F}+1.3\,\mathrm{meV}$. This observation implies that the maximum energy that can be transferred to equilibrium electrons in the detector circuit is $\Delta E^{\rm max}\equiv E_{\mathrm{kin}}-E_\mathrm{F}\simeq1.3\,\mathrm{meV}$. To further illustrate this exceptional behavior several $I_3$--$E_\mathrm{B3}$ traces at constant $V_\mathrm{SD}$ (indicated by horizontal lines in Fig.\ \ref{fig2}\textbf{a}) are plotted in Fig.\ \ref{fig2}\textbf{b}. The larger the injection energy $\left|eV_\mathrm{SD}\right|$ the sharper is the current onset at $E_\mathrm{B3}-E_\mathrm{F}\simeq\Delta E^{\rm max}$.

At low temperatures energy exchange between mesoscopic circuits is usually attributed to Coulomb interaction as indeed observed in Coulomb--drag experiments \cite{Gramila91,Sivan92,Lilly98,Tarucha06}. Here, in our experiments, the upper bound $\Delta E^{\rm max}$ of energy quanta transferred between emitter and detector reflects that the energy is mediated by interface acoustic phonons: Hot injected electrons can relax by emission of acoustic phonons \cite{Ridley91}. In contrast to electrons, acoustic phonons can pass the electrostatic barrier (B1) between emitter and detector circuits. Energy and momentum conservation restrict the emission of interface acoustic phonons by electrons with momentum $\hbar k_\mathrm{e}$ to momenta $k_\mathrm{ph}\lesssim 2\hbar k_\mathrm{e}$, corresponding to backscattered electrons in the 2DES. With the same consideration only interface acoustic phonons with $k_\mathrm{ph}\lesssim 2\hbar k_\mathrm{F}$ can be absorbed by equilibrium electrons in the detector. This situation is indicated in the inset of Fig.\ \ref{fig2}\textbf{b}, picturing the parabolic electron dispersion relation within the 2DES. The blue line indicates all possible states the electron (black circle), originally at the Fermi-energy, can be scattered into by absorption of an interface acoustic phonon. Thus scattered electrons drive the analyzer current in the detector circuit. With the upper bound $\Delta E^{\rm max}$ measured and the known Fermi momentum $\hbar k_\mathrm{F}$ in the 2DES we obtain with $\Delta E^{\rm max}\simeq E_{\rm ph}(2k_{\rm F})=2\hbar k_{\rm F}v_{\rm s}$ a sound velocity of $v_\mathrm{s}\simeq 6\,\mathrm{km/s}$, in good agreement with literature values of $v \simeq 5.3\,\mathrm{km/s}$ for longitudinal acoustic phonons propagating in bulk--GaAs in the $[110]$--direction \cite{Blakemore}, the orientation of our Hall--bar. Our experiments show conclusively that the analyzer current is caused by both energy and momentum imbalance of non--equilibrium electrons excited by absorption of interface acoustic phonons in the unbiased detector circuit.

With increasing $V_\mathrm{SD}$ high energy electrons can emit phonons with momenta exceeding by far $2\hbar k_\mathrm{F}$. However, momentum conservation requires that these phonons have a large momentum component perpendicular to the 2DES \cite{Karl88}. At low temperatures they propagate ballistically through the bulk crystal with a mean--free path beyond the crystal dimensions \cite{Karl88,Lehmann99}. As a consequence, only interface phonons are likely to be reabsorbed in the 2DES of the detector circuit and contribute to the analyzer current $I_3$. Correspondingly, the measured $I_3$ is typically five orders of magnitude smaller than the injector current $I_\mathrm{SD}$.

To avoid excessive power dissipation at large $\left|V_\mathrm{SD}\right|$, QPC1 is tuned to be highly resistive. For $|V_\mathrm{SD}|\lesssim8\,$mV QPC1 is even completely closed, $I_\mathrm{SD}$ vanishes, and therefore also $I_3$ (horizontal onset in Fig.\ \ref{fig2}\textbf{a}). In order to explore electron--phonon scattering at small energies we instead tune barrier B2 to be opaque for electrons and employ B1 as injector adjusted to a smaller resistance. The corresponding measurement is shown in Fig.\ \ref{fig2}\textbf{c} displaying the analyzer current $I_3$ as a function of $E_\mathrm{B3}-E_\mathrm{F}$ in the bias--range $-3\,\mathrm{mV}\le V_\mathrm{SD}\le3\,\mathrm{mV}$. Fig.\ \ref{fig2}\textbf{d} plots $I_3$--$V_\mathrm{SD}$ traces for constant $E_\mathrm{B3}$ (along the vertical lines in Fig.\ \ref{fig2}\textbf{c}). Also shown is the measured injector current $I_\mathrm{SD}$ versus $V_\mathrm{SD}$ (rhs axis). It forms a straight line reflecting that B1 acts as a constant resistance. Nevertheless, $I_3$ still vanishes for $\left|V_\mathrm{SD} \right| \lesssim 0.8\,\mathrm{mV}$, independent of the analyzer barrier height $E_\mathrm{B3}$ (Figs.\ \ref{fig2}\textbf{c} and \ref{fig2}\textbf{d}). Such a low--energy onset suggests that the interaction mechanism between emitter and detector strongly depends on energy. Note that energy transfer mediated by interface acoustic phonons is expected to strongly increase as their momenta approach $2\hbar k_\mathrm{F}$ (see inset in Fig.\ \ref{fig2}\textbf{b}) \cite{Ridley91,comment2}. Similar onsets have been observed in recent experiments on interacting mesoscopic circuits \cite{Khrapai06,Khrapai07}. No such onset behavior has been reported in experiments where the energy transfer between mesoscopic systems is mediated by potential fluctuations caused by moving charges \cite{Kouwenhoven06,Ensslin07}.

In Fig.\ \ref{fig2}\textbf{c} we find $I_3>0$ independent of the sign of $V_\mathrm{SD}$. Clearly, the detector circuit acts as a uni--directional current source, driven by phonons originating in the emitter. In the electrically separate detector electrons absorb such interface phonons predominantly close to the emitter. Then the excited electrons move in the direction of the transferred momentum towards barrier B3 where they can contribute to the analyzer current $I_3$. The latter is considerably smaller for $V_\mathrm{SD}>0$ compared to the case of $V_\mathrm{SD}<0$. We relate this to the initial momentum of the hot electrons in the emitter which is for $V_\mathrm{SD}>0$ directed away from the detector. In this case and in contrast to $V_\mathrm{SD}<0$ an additional scattering process is needed to reverse the momentum towards the detector. 

\begin{figure}[ht]
\centering
\includegraphics[angle=0]{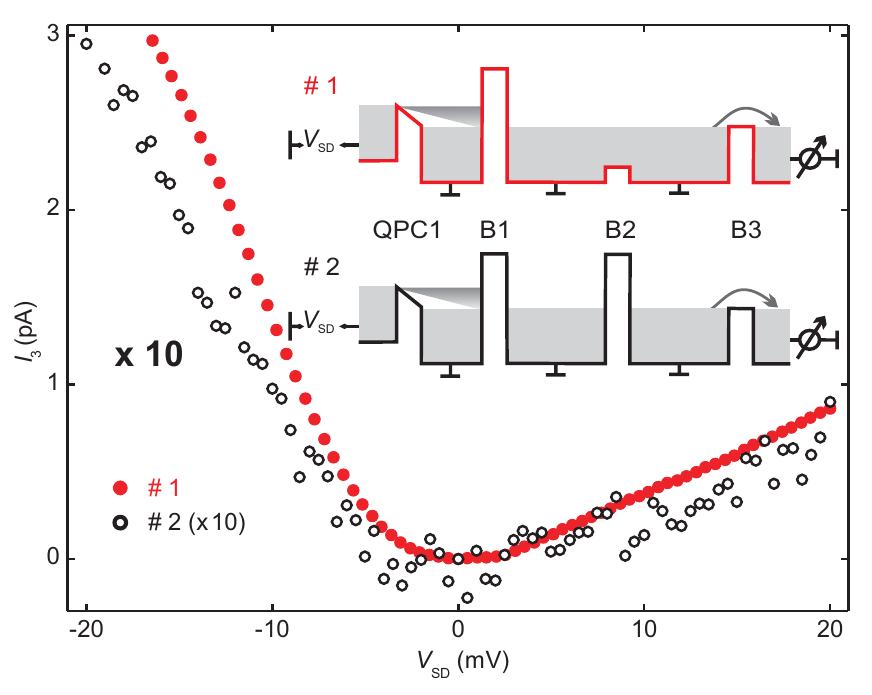}
\caption{\label{fig3} \textbf{Mean--free--path of interface acoustic phonons:} (color online) Analyzer current $I_3$ as a function of the emitter bias $V_\mathrm{SD}$ for the two experimental setups sketched in the inset. In setup \#1 B2 is left open (as for Fig.\ \ref{fig2}\textbf{a}) while in setup \#2 both barriers B1 and B2 are opaque for electrons. The analyzer barrier is tuned to $E_\mathrm{B3}\simeq E_\mathrm{F}$. In setup \#2 $I_3$ is reduced by about a factor of ten. Hence, the distance between B1 and B2 of about $1\mu$\,m roughly corresponds to the phonon mean--free--path.}
\end{figure}

Compared to elastic scattering of ballistic electrons at the Fermi--surface  \cite{Beenakker91} non--equilibrium interactions at higher energies remain a challenging subject. Hot electrons can relax their excess energy either via electron--electron scattering \cite{kaya07,Govorov2004,Predel2000,Chaplik71}, via electromagnetic fields generated by charge fluctuations \cite{Kouwenhoven06,Ensslin07,Gustavson08}, or via the emission of phonons \cite{Ridley91,Sivan89,Dzurak92}. Inelastic electron--phonon scattering in the 2DES for electrons with an excess energy of $\Delta E\simeq1\,\mathrm{meV}$ results in a mean--free path of $l_\mathrm{ep}\sim 100\,\mu$m \cite{Predel2000,Ridley91,Bockelmann90} considerably longer than the electron--electron scattering length of $l_\mathrm{ee}\sim 8\,\mu$m \cite{Quinn82}. Both length scales are longer than the elastic mean--free path of electrons, limited to $l_\mathrm{m}\sim1\,\mu$m by the geometric boundaries of the device. In Fig.\ \ref{fig3}
we investigate the length scale $l_\mathrm{pe}$ of the  reabsorption of interface acoustic phonons within the 2DES. We compare two different experimental situations as sketched in the inset of Fig.\ \ref{fig3}. Configuration \#1 is essentially identical to the one established in the experiment of Figs.\ \ref{fig2}\textbf{a} and \ref{fig2}\textbf{b} and displays the phonon--driven current as a function of $V_\mathrm{SD}$ with the analyzer barrier adjusted to $E_\mathrm{B3}\simeq E_\mathrm{F}$. In configuration \#2 an additional barrier (B2) is raised well above the Fermi level ($E_\mathrm{B2}\gg E_\mathrm{F}$). Now the resulting phonon--driven current is about a factor of ten smaller compared to configuration \#1 but exhibits almost the same dependence on $V_\mathrm{SD}$. This finding implies that most phonons passing B1 are reabsorbed by the 2DES before reaching barrier B2 and thus cannot contribute to the phonon--driven current. As the distance between barriers B1 and B2 is $1\,\mu$m we consider this as an upper limit for the interface phonon mean--free path $l_\mathrm{pe}$. The corresponding transition rates of $\sim (200\,\mathrm{ps})^{-1}$ agree roughly with theoretical estimates \cite{Ridley91,Bockelmann90} and $l_\mathrm{pe}/l_\mathrm{ep}$ is of the order of the ratio of sound and Fermi velocity, as expected.

In conclusion, our experiments on interacting non--equilibrium mesoscopic circuits underline the importance of energy transfer mediated via interface acoustic phonons and generated by ballistically moving electrons driven out of equilibrium. In particular, they demonstrate conclusively that this energy transfer between a non--equilibrium nanoscale circuit, serving as emitter, and an adjacent detector circuit in equilibrium is bounded by the energy of interface acoustic phonons with momentum $2\hbar k_\mathrm{F}$. This is the maximum momentum that can be transferred to equilibrium electrons under conservation of momentum and energy. Since such phonon--mediated interactions reduce the coherence times of quantum states in confined electron systems their study and understanding is important for the realization of semiconductor--based coherent quantum devices. Beyond we establish a method to spectroscopy interface acoustic phonons in a new regime up to momenta of $2\hbar k_{\rm F}$.

We thank A.\,O.\ Govorov, W.\ Dietsche, M.\ Heiblum, V.\,S.\ Khrapai, and K.\,F.\ Renk for stimulating discussions and D.\ Harbusch, D.\ Taubert, M.\ Kroner as well as S.\ Seidl for helpful comments. Financial support by the German Science Foundation via SFB 631 as well as the Germany Israel program DIP and by the German Excellence Initiative via the "Nanosystems Initiative Munich (NIM)" is gratefully acknowledged.


\begin{thebibliography}{apssamp}
\bibitem{comment1}Barriers B1, B2, B3 are calibrated with likewise results.
\bibitem{comment2}The 2DES screans the piezo--electric coupling between electrons and interface--phonons with small $k$--vectors. This screening effect hinders the emission and absorption of low energy interface phonons and might very well contribute to the observed low energy onset.
\end{thebibliography}
\end{document}